\documentclass[12pt]{article}

\usepackage{scicite}
\usepackage{times}

\usepackage{graphicx}
\usepackage{amsmath, amssymb, gensymb}
\usepackage{bm}
\usepackage{float}
\usepackage{color}
\usepackage[boxed]{algorithm2e}
\usepackage{upgreek}
\usepackage{multirow}

\graphicspath{{figures/}}

\newcommand{\norm}[1]{\left\lVert#1\right\rVert}

\def\deltares{\delta_{r}}

\def\<{\left\langle}
\def\>{\right\rangle}
\newcommand{\micron}{$\upmu$m}
\newcommand{\nph}{n_\textup{ph}}
\newcommand{\npixtheta}{N_{{\rm pix},\theta}}
\DeclareMathOperator*{\argmin}{arg\,min}
\newcommand{\vecpsi}[1]{\boldsymbol{\psi}_{#1}}
\newcommand{\vecx}{\boldsymbol{n}}
\newcommand{\vecy}{\boldsymbol{y}_{\theta,k}}
\newcommand{\matr}{\boldsymbol{R}_\theta}
\newcommand{\mats}{\boldsymbol{S}_{j}}

\newcommand{\matm}{\boldsymbol{M}_{\boldsymbol{n}, \theta, \Delta z}}
\newcommand{\mata}{\boldsymbol{A}_{\boldsymbol{n}, \theta, j}}
\newcommand{\varpsi}{\psi_{j, k}}
\newcommand{\funcf}{f(\boldsymbol{n}, \theta, k, \Delta z, d)}
\newcommand{\aor}{\sigma_{\rm AOR}}

\newenvironment{sciabstract}{%
\begin{quote} \bf}
{\end{quote}}

\title{Three dimensions, two microscopes, one code: automatic
  differentiation for x-ray nanotomography beyond the depth of focus limit}

\author
{Ming Du$^{1}$, Youssef S. G. Nashed$^{2}$, Saugat Kandel$^{3}$, \\
  Dog{\u a} G{\"u}rsoy$^{4}$, and Chris Jacobsen$^{4,5,6,\ast}$ \\
\normalsize{$^{1}$Department of Materials Science, Northwestern
  University, Evanston, IL 60208, USA} \\
\normalsize{$^{2}$Mathematics and Computer Science Division, Argonne
  National Laboratory, Lemont, Illinois 60439, USA} \\
\normalsize{$^{3}$Applied Physics Program, Northwestern University,
  Evanston, IL 60208, USA} \\
\normalsize{$^{4}$Advanced Photon Source, Argonne National Laboratory,
  Argonne, IL 60439, USA} \\
\normalsize{$^{5}$Department of Physics and Astronomy, Northwestern
  University, Evanston, IL 60208, USA} \\
\normalsize{$^{6}$Chemistry of Life Processes Institute, Northwestern
  University, Evanston, IL 60208, USA} \\
\\
\normalsize{$^\ast$To whom correspondence should be addressed; E-mail:
  cjacobsen@anl.gov}
}

\date{}

\begin{document} 




\maketitle


\begin{sciabstract}
 Conventional tomographic reconstruction algorithms assume that one
  has obtained pure projection images, involving no within-specimen
  diffraction effects nor multiple scattering.  Advances in x-ray
  nanotomography are leading towards the violation of these
  assumptions, by combining the high penetration power of x-rays which
  enables thick specimens to be imaged, with improved spatial
  resolution which decreases the depth of focus of the imaging system.
  We describe a reconstruction method where multiple scattering and
  diffraction effects in thick samples are modeled by multislice
  propagation, and the 3D object function is retrieved through
  iterative optimization.  We show that the same proposed method works
  for both full-field microscopy, and for coherent scanning techniques
  like ptychography. Our implementation utilizes the optimization toolbox 
  and the automatic differentiation capability of the open-source deep 
  learning package TensorFlow, which demonstrates a much straightforward 
  way to solve optimization problems in computational imaging, and 
  endows our program great flexibility and portability.
\end{sciabstract}

\section{Introduction}

Depending on the photon energy used, X rays are able to penetrate into
samples with a thickness ranging from micrometers to centimeters.  
At the same time, x-ray microscopes are beginning to be able to deliver images
with sub-10 nanometer spatial resolution
\cite{mimura_nphys_2010,chao_optexp_2012,mohacsi_scirep_2017,bajt_lsa_2018}.
However, combining these characteristics is complicated by the
fact that any imaging method with spatial resolution $\deltares$ has a
depth of focus DOF limit \cite{born_1999,wang_jmic_2000} of 
\begin{equation}
  \mbox{DOF} = \frac{2}{0.61^{2}} \frac{\deltares^{2}}{\lambda} 
  \simeq 5.4 \deltares\,\frac{\deltares}{\lambda}.
  \label{eqn:dof}
\end{equation}
This is straightforward to understand in lens-based imaging systems.
However, even when lensless imaging methods involving wavefront
recovery are employed, the depth of focus limit of Eq.~\ref{eqn:dof}
gives the axial distance over which features can be considered
to all lie within a common transverse plane before subsequent wavefield
propagation effects are taken into account.  That is,
Eq.~\ref{eqn:dof} represents the limit of validity of the pure
projection approximation, within which a depth-extended object can be
treated as producing a simple pure projection image when viewed from
one illumination direction.  For objects thicker than the depth of
focus (DOF) limit, one must instead account for wave propagation
effects within the specimen.  This will be especially important for
fully exploiting the dramatic increases in coherent x-ray flux that the
next generation of synchrotron light sources will provide
\cite{eriksson_jsr_2014}.

\begin{figure}[t]
  \centerline{\includegraphics[width=0.6\textwidth]{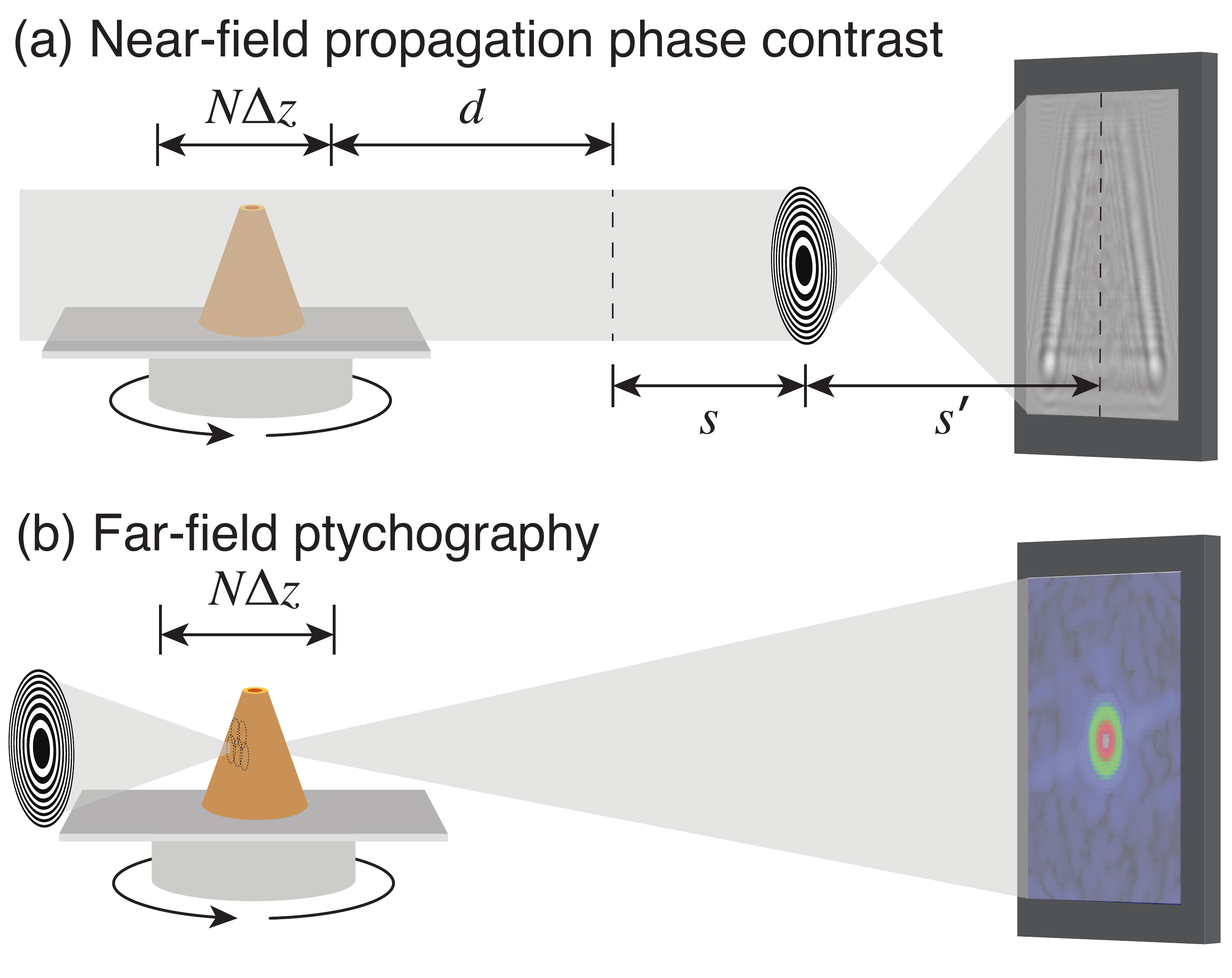}}
  \caption{Schematic representation of the two different microscope
    types used in our demonstration.  In the full-field mode (a),
    phase contrast is incorporated by allowing the wavefield leaving
    the object to undergo Fresnel propagation over a distance $d$
    before that plane is imaged by a lens (with
    $1/f=1/s+1/s^{\prime}$) onto a detector.  In ptychography, a small
    coherent beam spot (probe) is scanned through the object and the far-field
    diffraction intensities are recorded for each probe position.  The
    schematic here shows Fresnel zone plates as lenses for
    quasimonochromatic x-ray beams.}
  \label{fig:forward}
\end{figure}

\textcolor{black}{One approach to simulate wave propagation in a complex
object is the finite-difference method \cite{lars_opex_2017}. However,
due to the need to solve a series of partial differential equations,
the efficiency of this method relies on the availability of distributed
differential equation solvers, which are usually sophisticated to implement.
On the other hand,}
Multislice wave propagation \cite{cowley_actacryst_1957a} is a historic, simple 
but still powerful method allowing one to account for wave diffraction
in a inhomogeneous medium. The multislice simulation method 
subdivides the propagation problem into
a series of elemental modulation and propagation operations, 
and accounts for the change of the probe wave throughout the object instead
of assuming a constant probe. Hence, it can provide accurate numerical results propagation
through a complicated object, and remains valid over a much larger 
object thickness compared to diffraction tomography models assuming
single scattering \cite{devaney_optlett_1981, devaney_ui_1982}. 
The incorporation of multislice propagation could stand as a novel and
reliable strategy for the reconstruction of beyond-DOF objects.

We describe here an approach for imaging objects that extend beyond
the DOF limit, and within which multiple scattering might take place. 
We formulated the 3D object reconstruction problem as a minimization 
problem which incorporates a data fidelity term (L2-norm of simulated
and measured data) and a regularization term (L1-norm of the object 
and its gradient), where the multislice wave propagation is used to 
accurately model the exit wave leaving the object. Because the new 
model better captures the wave-object interactions for any object 
size, the same model can be applied to reconstruct either near-field 
imaging with propagation phase contrast or ptychography 
(Fig.~\ref{fig:forward}) without the 
need for any modification. We used the Adaptive Moment Estimation (Adam) 
optimizer that is implemented in TensorFlow, which is Google Brain’s 
open-source software library. The automatic differentiation capability 
in TensorFlow allows us to implement the optimization problems with 
minor tweaks. With this approach, we are able to use one computer 
code for two different types of microscopes to reconstruct 3D objects 
beyond the DOF limit. 

It is worthwhile noticing that while there exist several multislice-based
reconstruction method which have proven success in several imaging
scenarios \cite{maiden_josaa_2012,tsai_optexp_2016b,gilles_optica_2018}, 
our method differs from them in a few aspects.
Our implementation provides both a
ptychography mode and a full-field mode, while the above methods are
concerned with ptychography alone. In addition, instead of requiring
planes that are axially separated by 1 DOF of more, in our method the
spacing between slices can be equal to the lateral pixel width which
allows for an isotropic voxel size. Finally, the method for updating
the object function is different. In \cite{maiden_josaa_2012}, slices
are updated sequentially using an update function that resembles the
modulus replacement operation in ePIE, a reconstruction engine for 2D
ptychography. In \cite{tsai_optexp_2016b}, the first method described
is similarly based on modulus replacement, while the second method
involves the minimization of a loss function that has a similar form
of ours. However, in our approach to full-field microscopy, the loss
equation is constructed also with a sparsity constraint, and
non-negative and finite support constraints are applied to the object
function throughout the minimization process. This also distinguishes
our work with \cite{gilles_optica_2018}. Lastly, our employment
of automatic differentiation through the widely used software package
TensorFlow renders the implementation highly accessible and
flexible. On the top of the first reports of using AD in phase
retrieval problems
\cite{nashed_procedia_2017,ghosh_iccphot_2018,kandel_optexp_2019}, our
work reinforces the vast potential of AD for a large variety of
computational imaging tasks.

\section{Imaging beyond the depth-of-focus limit}

Present-day x-ray nanotomography is usually done within the depth of
focus limit of Eq.~\ref{eqn:dof} \cite{lovric_jac_2013, shahbazi_scirep_2018, 
yu_opex_2018, barrioberovila_materials_2017, krenkel_scirep_2016, 
mokso_apl_2007, lussani_rsi_2015, groso_opex_2006, hieber_scirep_2016}, 
such as with 1 \micron~resolution at
25 keV (giving $\lambda=0.050$ nm and DOF=110 mm), or 20 nm resolution
at 6.2 keV (giving $\lambda=0.20$ nm and DOF=11 $\mu$m).  In these
cases, one can obtain an image that represents a pure projection
through the specimen at each rotation angle by using standard phase 
retrieval methods based on the inversion of the Transport-of-Intensity 
equation \cite{burvall_optexp_2011}; one can then use standard
tomographic reconstruction algorithms such as filtered backprojection.
For objects that are thicker and/or interact more strongly, the
complete solution of the wave function of electromagnetic wave within
an inhomogeneous scattering potential field results in an recursive
equation. With the first iteration, one arrives at the first Born
approximation, which physically accounts for single scattering within
the sample. On this basis, one can approximate the
imaging of thicker specimens by acknowledging the fact that the
far-field diffraction pattern of an object provides information on the
surface of the Ewald sphere corresponding to the beam energy and
viewing direction \cite{wolf_oc_1969}. This re-mapping of Fourier space information from a
plane (pure projection), to the surface of the Ewald sphere, is used in
filtered backpropagation algorithms in diffraction tomography
\cite{devaney_ui_1982}.   It has been widely applied in tomographic
diffractive microscopy with visible light \cite{haeberle_jmo_2010} and
has been demonstrated in
x-ray coherent diffraction imaging \cite{chapman_josaa_2006}. 

One approach that has been developed for imaging beyond the DOF limit
is multislice ptychography \cite{maiden_josaa_2012}.  In standard
ptychography \cite{hoppe_aca1_1969,faulkner_prl_2004}, one scans a
finite sized coherent beam with overlap across a planar sample and
separates or factorizes the probe from the optical modulation at each
scan position.  Multislice ptychography is based on utilization of the
multislice method \cite{cowley_actacryst_1957a} (also known as the beam
propagation method \cite{vanroey_josa_1981}) to propagate a beam
through a thick object, where the refractive effect of the first thin
slab of the object is applied to the incident wavefield, the wavefield
is free-space-propagated to the next slab position, and the process is
repeated until one obtains the exit wave leaving the object (which can
then be free-space propagated to a far-field detector, for example).
If the object is in fact comprised of a series of discrete planes
separated axially by 1 DOF or more, one can factorize the probe from
both transverse positions, and axial planes as well. One can also
account for violation of the Born approximation, in that the
object-modulated exit wave from the upstream plane is propagated to
the next axial plane in a recursive manner through all planes.  This
approach has been used with success in ptychography using visible
light \cite{godden_optexp_2014}, X rays
\cite{suzuki_prl_2014,tsai_optexp_2016b,ozturk_optica_2018}, and
electrons \cite{gao_natcomm_2017}.  It has also been used for
tomographic imaging of more continuous specimens by assuming that the
object could be represented by discrete axial planes separated by the
DOF \cite{li_scirep_2018}.  However, this assumption is only
approximately true, since one can often see image contrast variations
with defocus settings of less than 1 DOF (or the separation of the slices), 
especially in phase contrast
which is the dominant contrast mechanism in transmission x-ray
microscopy \cite{schmahl_xrmtaiwan}. In that case, variation of
features along the beam axis between each two adjacent slices will not be
captured. In addition, multislice
ptychographic tomography as implemented above requires
phase-unwrapping of individual \cite{dierolf_nature_2010} or the
summation \cite{li_scirep_2018} of phase contrast images obtained
prior to their use in tomographic reconstruction, and this phase
unwrapping process can sometimes present difficulties or inaccuracies.

Therefore it can be advantageous to use reconstruction methods that
use a forward model of multislice propagation in a continuous object 
and retrieve directly
the refractive indices of the objects instead of the phase of the
exiting waves, so that
multiple scattering effects are included and no phase
unwrapping is required.  Calculations for x-ray imaging of biological
specimens show that one must begin to account for multiple scattering
effects at a specimen thickness of a few $\mu$m in soft x-ray imaging
at 0.5 keV, and a few tens of $\mu$m in hard x-ray imaging at 15 keV
\cite{du_ultramic_2018}. The need for the
inclusion of multiple scattering effects is well known in optical
diffraction microscopy \cite{haeberle_jmo_2010}, and we have
shown that this approach can be used for x-ray microscopy 
as well \cite{gilles_optica_2018}.  What we describe below is
a new approach that is different than the above: it uses
the method of automatic differentiation 
to carry out the reconstruction, which offers greater flexibility on imaging method and
for incorporating various constraints on the object as numerical optimization regularizers.

\section{Formulation of the image reconstruction problem}

Our approach is to treat image reconstruction of objects beyond the
DOF limit as an numerical optimization problem.  That is, we wish to find the
optimal parameter set $\vecx_0$ of the forward model $f$
by minimizing an objective function $L$, leading to a
solution of
\begin{equation}
  \vecx_0 = \argmin_{\vecx} L[f(\vecx), \boldsymbol{y}],
  \\ \text{~subject to } \vecx \in \Phi
  \label{eqn:minimized_cost}
\end{equation}
where the observable $\boldsymbol{y}$ is the set of experimental measurements
(near-field images or far-field diffraction patterns), and $\Phi$
is the manifold of contraints that $\vecx$ is subject to.  The parameter
set $\vecx$ will be defined in Eq.~\ref{eqn:x_equation} below
to be proportional to the x-ray refractive index (RI) distribution within the
object's voxel grid positions $\boldsymbol{r}$. This refractive
index is written as
\begin{equation}
  n(\boldsymbol{r}) = 1 - \delta(\boldsymbol{r}) -  i\beta(\boldsymbol{r}).
  \label{eqn:refractive_index}
\end{equation}
where the values of $\delta$ and $\beta$ for various materials are
readily obtained from tabulations \cite{henke_adndt_1993}.  Except at
photon energies right near x-ray absorption edges where anomalous
dispersion effects can appear, $\delta$ and $\beta$ have small
positive values (typically $\delta\simeq 10^{-4}$ and
$\beta \simeq 10^{-5}$ to $10^{-6}$) so a positivity constraint can be
applied to their solution.  One can also apply a sparsity constraint
for objects that are relatively discrete in space
\cite{kamilov_ieeetci_2016}, and in most cases tomography experiments
are designed so that the object fits within the field of view so one
can also apply a finite support constraint on the solution of
$n(\boldsymbol{r})$.

Because the size of the optimization problem is high,
efficient mechanisms must be used to find gradients for each iteration 
of the first-order solver used in optimizing Eq.~\ref{eqn:minimized_cost}.  
If one is always considering one type of
imaging experiment, one can calculate derivatives of the cost function
and indeed this approach has been used with success for simulations of
x-ray ptychographic reconstruction of objects beyond the DOF limit
\cite{gilles_optica_2018}.  However, if one wishes to be able to treat
multiple imaging methods (so as to compare or benchmark their properties and performances, for
example) and include a variety of regularizers, other approaches that
place the burden of finding minimization strategies on a computer
rather than a scientist can have advantages.  One
approach is to represent multislice propagation with a computational
architecture resembling a convolutional
neural network and use mathematical formulations that are common in
machine learning to solve for the object
that matches the observations, as has been demonstrated for
diffraction microscopy using visible light \cite{kamilov_optica_2015}.
Automatic differentiation (AD) \cite{rall_1981} provides another
approach which was suggested for use in phase retrieval problems
\cite{jurling_josaa_2014}, and then successfully implemented for x-ray
ptychography \cite{nashed_procedia_2017,ghosh_iccphot_2018}.  More recently, the
adaptability of automatic differentiation to a variety of coherent
diffraction imaging methods has been demonstrated
\cite{kandel_optexp_2019}, and a variety of software toolkits are now
available to utilize this method (spurred on by their use for
constructing the trainer module in supervised machine learning
programs).  We use this AD approach to reconstruct
beyond-depth-of-field imaging in two successful imaging methods, and
thus gain insight on their relative advantages and complications.

As noted above, in x-ray ptychography one scans a finite sized
coherent beam through a series $k$ of overlapping probe positions
across the specimen, and collects the far-field diffraction pattern
from each.  Because the extent of the far-field diffraction pattern is
determined by the scattering properties of the object rather than the
spatial resolution of the probe, one can obtain reconstructed x-ray
images with a spatial resolution far finer than the size of the probe
\cite{rodenburg_prl_2007,thibault_science_2008}.  In contrast, point
projection x-ray microscopy \cite{cosslett_nature_1951}, where an
object is placed downstream of a point source of radiation, provides
geometric magnification of the object with a penumbral blur limit
given by the source size, plus diffraction blurring that can be
compensated for by near-field wave backpropagation
\cite{cloetens_apl_1999,mokso_apl_2007}.  Because near-field
diffraction blurring is localized to a region given by the finest
reconstructed feature size times the propagation distance divided by
the wavelength, one can make use of illumination with a coherence
width equal to this region rather than the entire illumination field
as required for ptychography, and thus make more complete use of
partially coherent sources.  Since the advancing of x-ray imaging 
instruments has granted considerable potential to both techniques
in their imaging applications for thick samples with high resolution, 
it is of interest to understand
the beyond-DOF imaging properties of both of these approaches.

\section{Algorithm implementation}

\subsection{The forward model}

\begin{figure}
	\centering
	\includegraphics[width=0.7\textwidth]{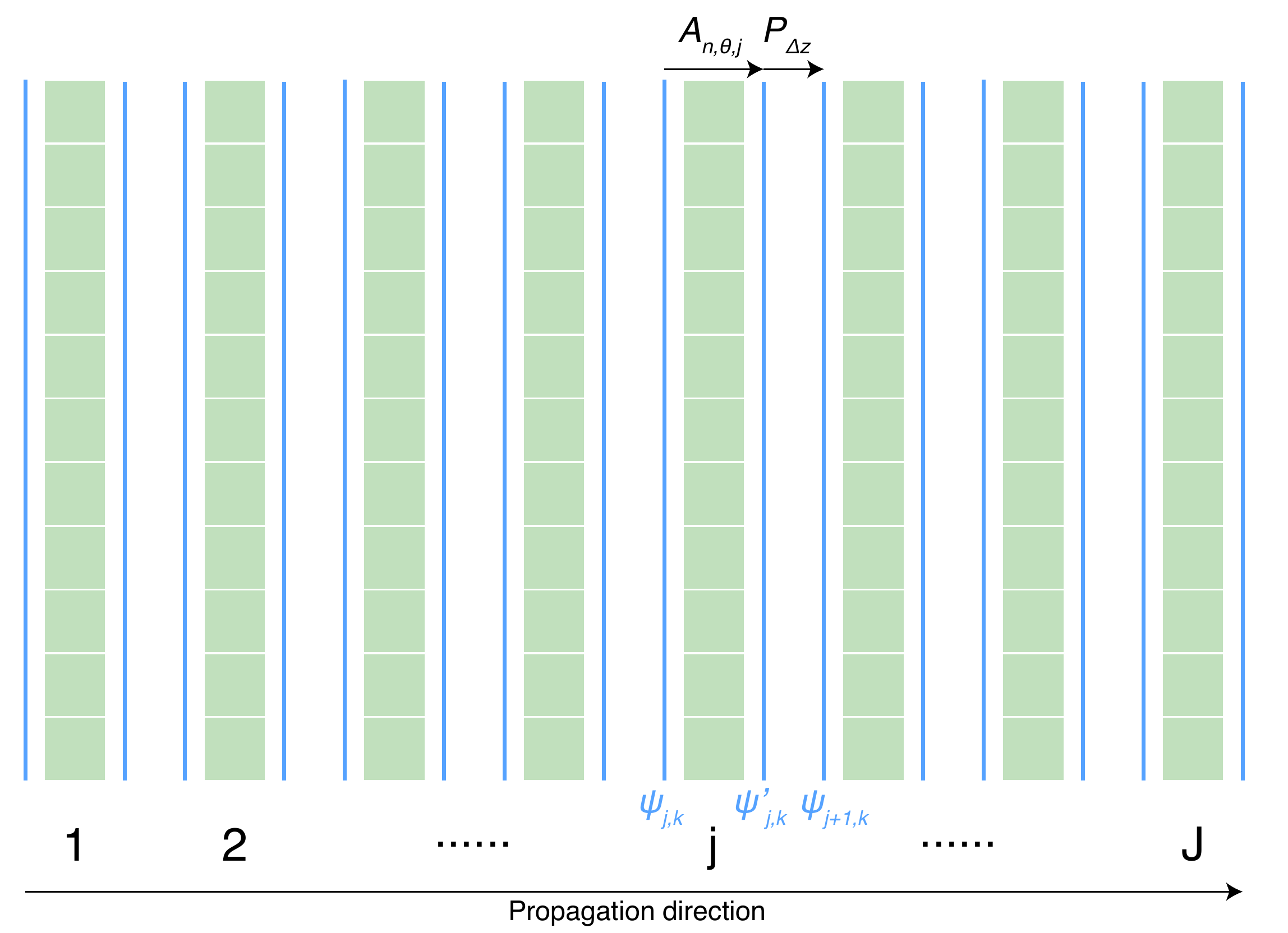}
	\caption{Illustration of multislice propagation.}
	\label{fig:multislice}
\end{figure}

We use the multislice method
\cite{cowley_actacryst_1957a} to calculate the wave exiting the
object, since it incorporates multiple scattering while also
accounting for effects such as waveguide phenomena
\cite{li_optexp_2017} (with the sole limitation of ignoring
backscattering which is neglible for all cases except Bragg
diffraction from perfect crystals or synthetic multilayers).  
As illustrated in Fig.~\ref{fig:multislice}, in multislice propagation, the object is divided into $J$ slices
along the beam axis. 
The wavefield $\varpsi(x,y,z_{j})$ from probe position $k$ 
(for full-field, $k=0$ for the first and only
probe position) entering the
$j^{\rm th}$ slice with thickness $\Delta z$ is modulated by the slice
to yield a wavefield $\varpsi^{\prime}(x,y,z_{j})$ of
\begin{align}
\begin{split}
  \varpsi^{\prime}(x, y,z_{j}) &= \varpsi(x, y,z_{i})
  \exp\Bigl[-\frac{2\pi \Delta z}{\lambda}
  i[1 - \delta(x, y, z_{j}) - i\beta(x, y, z_{j})]\Bigr] \\
  &= \varpsi(x, y,z_{i})\exp\Bigl(-\frac{2\pi \Delta z}{\lambda}i\Bigr)\exp(n_j)
  \label{eqn:multislice_modulation}
\end{split}
\end{align}

with
\begin{equation}
  n_j = \frac{2\pi (\Delta z)}{\lambda}
  [i\delta(x, y, z_{j}) - \beta(x, y, z_{j})].
  \label{eqn:x_equation}
\end{equation}
We will denote the RI distribution of the entire object
by vector $\vecx$ in the following text. 

The wavefront is then free-space propagated to the next
slice according to the Fresnel diffraction integral given by
\begin{equation}
  \psi_{j+1,k}(x, y,z_{j+1}) = \varpsi^{\prime}(x, y,z_{j}) * h_{\Delta z}(x, y)
  \label{eqn:propagation}
\end{equation}
where $*$ denotes the convolution operator, and $h_{\Delta z}(x, y)$ is
the Fresnel propagator given by
\begin{equation}
  h_{\Delta z}(x, y) = \exp\Bigl[-i\frac{\pi}{\lambda \Delta z}
  (x^{2} +y^{2})\Bigr].
  \label{eqn:propagator_function}
\end{equation}
This process is repeated for all $J$ slices until one obtains the exit
wave leaving the object.  
Here the slice thickness $\Delta z$ can be equal to the transverse pixel
size $\Delta x$, or for computational speed one can combine multiple
slices from the 3D volume together provided one satisfies the
condition of
\begin{equation}
  \Delta z \le \frac{2 \tilde{n} Q}{\pi} \frac{(\Delta x)^{2}}{\lambda} 
  \label{eqn:slice_thickness_klein}
\end{equation}
where $\tilde{n}$ is the mean RI, $\Delta x$ is the pixel size,
and $Q$ is the
Klein-Cook parameter for which values of $Q \lesssim 1$ represent the
case of plane grating rather than volume grating diffraction
\cite{klein_ieeetsu_1967}.  

In fact, multislice propagation is only one of several steps that must
be combined in the forward model of tomography beyond the DOF limit.
Tomography acquisition requires the illumination of the object from 
multiple rotation angles $\theta$. Our approach to account for the rotation 
is to rotate the object onto a constant
wave propagation direction, rather than to rotate the illumination;
this is done with a rotation operator $\boldsymbol{R}_{\theta}$.
After carrying out the sequence of $J$ multislice propagation steps
through the rotated object, we then need to apply the operator
$\boldsymbol{P}_d$ to take the exit wave $\vecpsi{J,k}$ from the
object to the plane of the detector, either using free space
propagation as described by $\psi_{J,k} * h(x,y,d)$ for near-field
propagation of distance $d$, or a simple Fourier transform for
far-field propagation (in the Fraunhofer approximation). This leads us
to a combined forward operation of
\begin{equation}
  \funcf = \boldsymbol{P}_{d}\matm\vecpsi{0,k}.
  \label{eqn:combined_forward_operations}
\end{equation}
In Eq. \ref{eqn:combined_forward_operations}, $\matm$ 
is the multislice propagation operator which is a function of the 3D object 
and the incident probe. $\matm$ describes the exit wave that leaves the depth-extended specimen.
It can be compactly written as
\begin{equation}
    \matm = \prod_j^J \boldsymbol{P}_{\Delta z}\mata
    \label{eqn:multislice_operator}
\end{equation}
with
\begin{equation}
	\mata =
        \exp[\mbox{diag}(\mats\boldsymbol{R}_\theta{\vecx})]
        \label{eqn:a_diag}
\end{equation}
where $\mats$ is a matrix that samples the $j$-th slice of 
column vector $\boldsymbol{R}_\theta{\vecx}$. 
If the total number of voxels of the object function is $N_v$,
and the number of pixels in the detector is $N_p$, then
$\vecx$ is a $N_v \times 1$ column vector, $\matr$ is a $N_v \times N_v$
square matrix, and $\mats$ is in the shape of $N_p \times N_v$, so that
it yields an $N_p \times 1$ column vector, which is of the same size
as the wavefront $\vecpsi{0,k}$. Multiplying diagonal
matrix $\mata$ with the wavefront vector $\vecpsi{0,k}$
is exactly the wavefront modulation given in Eq. \ref{eqn:multislice_modulation}.

\subsection{Constrained loss function minimization}

It has been pointed out for conventional 2D coherent diffraction imaging that
since only magnitude information is available in the detected far-field 
diffraction pattern, a reconstructable object should be spatially isolated,
with prior knowledge about the geometry of the object incorporated into the
reconstruction through zeroing out pixels out of the object boundaries. 
This is known as a ``finite support'' constraint \cite{fienup_optlett_1978}. 
Ptychography does not require a finite
support constraint applied in object space because the bounded probe 
itself is already a form of finite support constraint, and furthermore the overlap
between adjacent probe positions supplies sufficient information to solve for
all object unknowns. 

In our work where a 3D object is retrieved, the same criteria are
followed.  As will be shown in the results section, the ptychography
reconstruction of the object using our algorithm does not need prior
knowledge about the spatial extent of the sample.  However, a finite
support constraint was found to be necessary for the full-field case.  The initial
finite support mask is determined following the procedures
below. First, single-distance near-field phase retrieval
\cite{paganin_jmic_2002} is applied to all projection images to obtain
a first guess of the weak phase contrast projection through the
object.  We then use the tomographic set of these projections to
obtain a rough guess of the object support using the standard filtered
backprojection tomographic reconstruction algorithm.  The reconstructed
volume is then Gaussian filtered to remove noise and local
discontinuities. A Boolean mask is subsequently obtained by
thresholding the filtered object, which yields a support mask denoted
by set $\Theta$.  During the iterative reconstruction process, the
finite support is contracted to exclude low-value pixels for every
epoch, a technique known as ``shrink-wrap'' in conventional CDI
processing \cite{marchesini_prb_2003}.

We then wish to compare the present guess of the detected amplitude
of $|\funcf|$ against the amplitude $\sqrt{\vecy}$ measured in the
experiment, and minimize the difference between the two as expressed
in a cost function of
\begin{equation}
  L = \frac{1}{N_\theta N_p N_k}\sum_{\theta,k}
          \norm{|\funcf| - \sqrt{\vecy}}_2^2 +
        \alpha_\delta|\vecx_\delta|_1 +
        \alpha_\beta|\vecx_\beta|_1 +
        \gamma\mbox{TV}(\vecx_\delta).
    \label{eqn:loss_func}
\end{equation}
That is, the solution of the object function should be given by
\begin{eqnarray}
  \vecx_0 &=& \argmin_{\vecx}(L) \\
    \nonumber & & \textup{subject to } 
      n_w = 0 \text{ for } n_w \not\in \Theta 
      \text{ and } n_w \ge 0 \text{ for } n_w \in \Theta.
    \label{eqn:loss_func}
\end{eqnarray}

Here, $n_w$ is the $w$-th element in $\vecx$, $\vecy$ is the
measured intensity at orientation angle $\theta$, $N_\theta$ is the
number of projection angles, $N_p$ is the number of pixels in each
$\vecy$, $N_k$ is the number of probe positions for each projection
angle (equals 1 for full-field), and $\alpha_\delta$ and
$\alpha_\beta$ are the scalar normalizing coefficients added to the L1-norm
regularization terms for the $\delta$-part and $\beta$-part of
$\vecx$ respectively. The separated regularization for the
two parts of the object is necessary since $\delta(\boldsymbol{r})$
and $\beta(\boldsymbol{r})$ of the same material typically differ by a
few orders of maginitudes.  Together, these two L1-norm terms enhance
the sparsity of the object function, which is useful when the object
is spatially discrete or contains a lot of empty space (such as a dispersion
of cells, or a hollow structure).  Finally, the anisotropic total
variation $\mbox{TV}(\vecx)$, weighted by coefficient
$\gamma$, enhances the sparsity of the gradient of the object
function, which suppresses noises and unwanted heterogeneities
\cite{grasmair_amo_2010}. This regularizer is expressed as
\begin{equation}
	\mbox{TV}(\vecx) = \sum_{l,m,n} [|x_{l+1,m,n} - x_{l,m,n}| 
	+ |x_{l,m+1,n} - x_{l,m,n}|
	+ |x_{l,m,n+1} - x_{l,m,n}|]
	\label{eqn:tv}
\end{equation}
where $l$, $m$, and $n$ are indices along the three axes of
$\vecx$. The TV regularizer is only applied to
$\vecx_\delta$ because it usually carriers higher contrast
and better structural information than $\vecx_\beta$ when
hard X rays are used.

The successful retrieval of
$\vecx$ requires the simultaneous pixel-wise update of it, guided by 
$\nabla_{\vecx} L$, which is the gradient of loss function $L$ in the 
parameter space of $\vecx$. 
%
We use TensorFlow \cite{tensorflow_2016}, a deep learning package
first initiated by Google but now available as an open-source toolkit,
for carrying out our AD reconstruction.  It provides a user-friendly
Python application programming interface (API), and the ability to
write a reconstruction code of relative simplicity and with easy
implementation on a variety of computing platforms.  The AD algorithm
uses the so-called ``back-propagation'' method to derive the partial
derivatives in a semi-analytical fashion \cite{rall_1981}. 
Here, the loss function $L$ is first evaluated
in the forward direction using Eq.~\ref{eqn:loss_func}, during which
the intermediate variables produced by every algebraic operation are
computed and stored. After that, the algorithm calculates the
derivative of $L$ with regards to the intermediate variables
immediately before $L$ using the values saved in memory. This is
repeated back through the entire computation model, and the gradient
of $L$ with regards to $\vecx$, $\nabla_{\vecx} L$,
is then found based on the chain rule of differentiation.  Compared to
symbolic differentiation which attempts to acquire the closed-form
expression of $\nabla_{\vecx} L$ before doing any numerical
calculation, AD is free from the problem of expression swell when the
forward model is complicated.  On the other hand, AD is also more
accurate than the finite difference method, which approximates
$\nabla_{\vecx} L = \Big(\partial f/\partial x_1, ...,
\partial f/\partial x_n\Big)$ with
$\partial f/\partial x_i \approx [f(\vecx +
h\boldsymbol{e}_i) - f(\vecx)]/h$ for a small $h$
\cite{griewank_2008}.  We use the well-established optimization
algorithm ADAM (ADAptive Moment estimation) to update $\vecx$
\cite{kingma_iclr_2015}. A brief description of the algorithm is
provided in the supplementary material.

\subsection{Computational performance enhancements}

Like conventional tomography, the dataset acquired for reconstruction
would generally involve a large number of rotation angles $N_\theta$
(though in multislice reconstruction methods one can reduce
$N_{\theta}$ from below the number one would have expected from the
Crowther criterion \cite{jacobsen_optlett_2018}).  The large value of
$N_{\theta}$ can lead to the use of considerable computation power in
the iterative update of $\vecx$.  To reduce this, we note
that the first term in Eq.~\ref{eqn:loss_func} is essentially an
expectation value of error per pixel, and it can be adequately
approximated by calculating the error over a subset of
$N_{\theta}$. This technique, known as ``minibatching'' 
(or ``ordered-subsets'' in tomography literature \cite{hudson_ieee_1994}), 
can speed up
the convergence of the algorithm by several times. For each minibatch,
the subset of $N_{\theta}$ to be processed is chosen randomly without
replacement, so that the entire collection of $N_{\theta}$ will be
fully gone through after a certain number of minibatches are
completed. We hereafter refer to this process as an ``epoch,'' using
terminology drawn from the machine learning community.

In an actual experiment, the presence of noise typically induces
uncertainty in the ``sub-loss function'' obtained from each
minibatch. In this case, a true global minimum is ill-defined, which
causes $\vecx$ to dangle 
at the late stage of the optimization and thus prevents a stable
convergence. For this reason, users of TensorFlow have the option to
aggregate the gradients calculated from several minibatches, and
apply them to the optimizer after the completion of all of these
minibatches. The larger sample amount for gradient calculation reduces
the statistical fluctuation induced by noise and guarantees a more
stable solution.

A multiscale technique is also used in this work to further improve
reconstruction speed and accuracy. While the algorithm requires that
$\vecx$ has the same number of lateral pixels as the measured
data, instead of directly reconstructing the object with the same
pixel size as the acquired projections, both $\vecx$ and
$\vecy$ are downsampled by a factor of $2^m$, where $m$ is an integer
so that the lateral dimension of $\vecx$ is not larger than
64 pixels.  The voxel size of the object is thus accordingly enlarged
by $2^m$ times.  A first pass reconstruction of $\vecx$ is
therefore computed rapidly.  The result is upsampled by a factor of 2,
and then used as the initial guess for the next pass, where the scale
of the object and projections are doubled. This process is repeated
until the object is reconstructed with the acquired voxel size.  By
initializing $\vecx$ with the result from a coarse pass for a
higher-resolution pass, the optimization begins at a location closer
to the global minimum in the parameter space of $L$, so that fewer
iterations are required to converge. This may also reduce the chance for
the optimizer to get trapped in a remote local minimum.

%

\subsection{Parallelization}

In view of the huge number of unknowns ($3.4\times 10^7$ for a $256^3$ object
because of the presence of both the $\delta$ and $\beta$ parts of the RI) 
to be solved in our algorithm,
parallelized computation is necessary to guarantee a reasonable
computational walltime (within a few hours for a $256^3$ object). 
We use a TensorFlow add-on called Horovod
\cite{horovod} to implement distributed parallelization using the
message passing interface (MPI) standard for inter-rank (or inter-process)
communication, rather than the TCP/IP protocol used by native
TensorFlow (MPI is faster on tightly bound high performance computer
clusters). Each thread (or worker) initializes and keeps its own
object function, and process a minibatch simultaneously with other
threads. When a rank finishes its minibatch, it waits for other
threads to finish theirs, after which the gradients obtained by all
threads are averaged. The averaged gradient is then used to update the
object functions in all threads. Since the volume of samples used for
gradient calculation is effectively enlarged by a factor that equals
the number of threads $n_{\textup{thrds}}$, this means that the actual
learning rate \cite{kingma_iclr_2015} for the ADAM optimizer is multiplied by
$n_{\textup{thrds}}$.

The code used in this work has been made publicly available on GitHub
in the repository named ``Adorym''\footnote{https://github.com/mdw771/adorym.},
which is an acronym for ``\textbf{A}utomatic
\textbf{D}ifferentiation-based \textbf{O}bject \textbf{R}etrieval with
d\textbf{Y}namic \textbf{M}odeling.''

\subsection{Reference reconstructions}

In order to compare the outcomes of the proposed algorithm with
methods that are conventionally used for phase retrieval, the
full-field data demonstrated in this work are also processed and
reconstructed by first performing an iterative 2D phase retrieval
method termed error reduction (ER; widely used in coherent diffraction
imaging \cite{fienup_optlett_1978}) for every projection image. The
workflow of ER can be summarized as follows:
\begin{enumerate}
\item Propagate the initial guess of the exit wavefront $\psi_0$ to
  the detector plane as
  \begin{equation*}
    \psi_0' = \boldsymbol{P}_d \psi_0;
  \end{equation*}	
\item Replace the magnitude of the wavefront with the modulus of the
  measured intensity $I_\theta$ as
  \begin{equation*}
    \psi_0'' = \frac{\psi_0'}{|\psi_0'|} \sqrt{I_\theta};
  \end{equation*}	
\item Backpropagate the wavefront to the exiting plane as
  \begin{equation*}
    \psi_0''' = \boldsymbol{P}_{-d} \psi_0'';
  \end{equation*}	
\item Mask out the pixels of the wavefront that do not belong to the
  finite support $\Theta_2$ by doing
  \begin{equation*}
    \psi_1 = \Big\{
    \begin{array}{lr}
      \psi_0'''(\boldsymbol{r}), & \boldsymbol{r} \in \Theta_2 \\
      0, & \boldsymbol{r} \not\in \Theta_2
    \end{array}.
  \end{equation*}
\item The above processes are then repeated until the mean square
  error between the calculated intensity and the measured intensity
  converges.

\end{enumerate}
The filtered backprojection or FBP tomographic reconstruction
method is then applied to phase-retrieved images to obtain a 3D
reconstruction result. This ER+FBP approach will be subsequently referred to
as ``pure projection tomography.''

\section{Computational experiment}

Our Adorym approach was tested in simulations of thick objects that
would normally involve multiple scattering along the beam path.  Two
virtual samples were investigated: one that is described below, and a
protein sample that is presented in Supplementary Material.  The
reconstructions of the silicon cone sample shown here were performed on
the computing cluster Cooley at the Argonne Leadership Computing
Facility.  Each node of this cluster is equipped with two 2.4 GHz
Intel Haswell E5-2620 v3 CPUs (12 cores total) and 384 GB RAM.
Due to the limit of memory, we did not use GPU acceleration on this machine.
TensorFlow version 1.4.0 was used as the computational engine for our
routine. For the ADAM optimizer built with TensorFlow, we used a step
size of $\lambda = 10^{-7}$, and first and second moment exponential
decay rates of $\beta_1 = 0.9$ and $\beta_2 = 0.999$, respectively.

\begin{figure}[tbp]
  \centerline{\includegraphics[height=0.6\textheight]{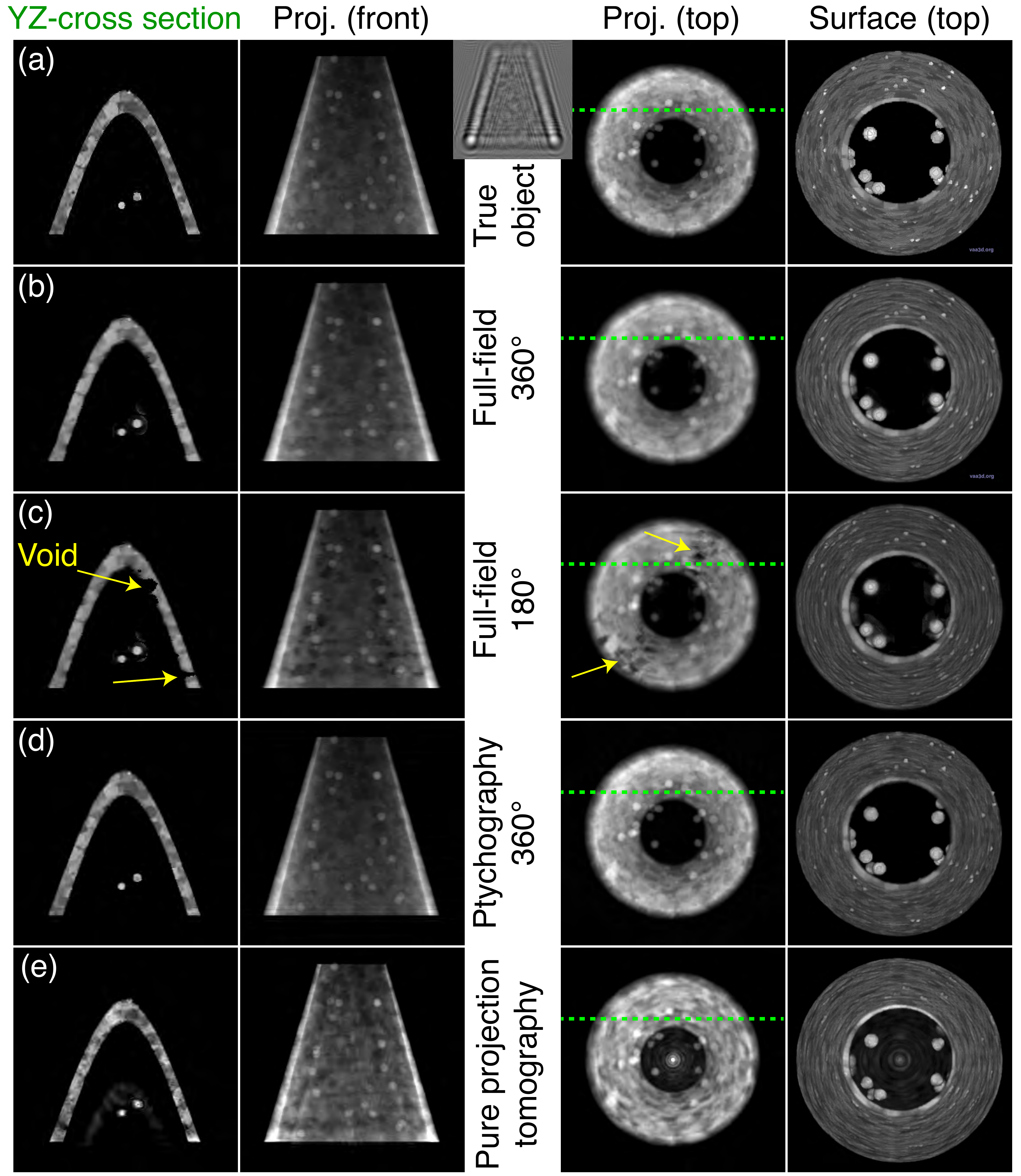}}
  \caption{Cone object used for computational experiments in beyond
    depth-of-focus x-ray imaging. The inset between column 2 and
    3 shows a near-field diffraction image used in full-field reconstruction. 
    The object (top row) and various
    reconstructed images (subsequent rows) are shown, displaying only
    the phase-shifting part $\delta$ of the x-ray RI
    since it provides higher contrast than the absorptive part
    $\beta$.  In keeping with the convention of most synchrotron
    tomography experiments, the object is assumed to be rotated around
    the vertical or $y$ axis.  The first column shows a single-plane
    section of the object in the $yz$ plane, with the location of the
    section shown by the green dashed line in the third column.  The
    second column shows a projection or summation through the 3D grid
    in that same direction, while the third column shows a projection
    or summation through the 3D grid viewed from above.  The fourth
    column shows the surface of the object, as rendered using Vaa3D
    \cite{vaa3d}.  Both the full-field (b) and the ptychography (d)
    reconstructions of data acquired over a 360$\degree{}$ rotation
    angle range reproduce the object with high fidelity.  In the
    full-field reconstruction from 180$\degree{}$ rotation data (c),
    voids appear where there is supposed to be material within the
    cone (in this case the illumination was incident from top, to
    right, to bottom in the perspective of the third and fourth
    columns, though the same type of behavior was observed with
    different 180$\degree{}$ illumination ranges).  In
    the pure projection tomographic reconstruction shown in the bottom
    row (e), one obtains an image with lower resolution and greater
    differences from the true object: the fine TiO$_{2}$ spheres on
    the outside of the cone object are blurred out, and the
    reconstruction shows spurious material inside the cone.  These
    images are all of a 256$\times$256 nm$^{2}$ field of view.}
  \label{fig:cone_results}
\end{figure}

We carried out computational experiments on a cone object created
using the optical simulation package \textit{XDesign}
\cite{ching_jsr_2017}.  In order to exceed the depth of focus limit
within a moderate array size, we chose to create a $(256)^{3}$ voxel
grid with 1 nm voxel size, and to use 5 keV x-ray beam energy.  As a
result, the x-ray wavelength was 0.248 nm and the depth of focus given
by Eq.~\ref{eqn:dof} was DOF=21.8 nm.  (Present-day x-ray microscopes
achieve a spatial resolution of more typically 15--30 nm as noted in
the introduction, but 1 nm represents a goal for the future).
Therefore the reconstruction grid was almost 12 times larger than the
depth of focus, so the object significantly excedes the depth-of-focus
limit. Within this grid, a hollow cone of silicon was
computationally created so as to resemble a thin-walled capillary
heated and then pulled.  The tube has a top diameter of 80 nm and a
bottom diameter of 200 nm, so that neither end fits within the DOF
limit.  To examine the algorithm's capability to restore fine details,
we also placed 50 TiO$_2$ nanospheres with radii ranging from 2 to 4
nm on the outer wall of the tube, as well as 10 larger spheres (5--13
nm in radius) inside the tube. The refractive indices of both
materials were generated using the open-source database package
\textit{Xraylib} \cite{xraylib}.  In addition, we added spherical
bubbles, or ``grains,'' whose refractive indices fluctuate within 30\%
of the original material, into the cone's body as a means to test the
ability of the algorithm to retrieve internal structure.  These grains
also served as marks in assessing the influence of photon noise on
ptychography results as will be discussed below.  The phase shifting
part $\delta$ of the x-ray RI for this
computationally-created object is shown in the top row of
Fig.~\ref{fig:cone_results}.

In x-ray full-field imaging, a variety of methods are available to
obtain a phase contrast image from detected intensities
\cite{wilkins_ptrsa_2014}, but propagation-based phase contrast is the
simplest to achieve experimentally.  We therefore assumed that the
x-ray wavefield exiting the object propagated downstream by a
sample-detector distance $d$ of 1 \micron~before the resulting wavefield
was imaged without loss by a 1 nm resolution lens onto a detector, as
shown in Fig.~\ref{fig:forward}.  This is beyond the state of the art
with present-day x-ray optics, but as noted above we chose parameters
so as to significantly exceed the depth of focus limit within a small
array size.  Within this optical configuration, we simulated the
recording of 500 images over a single-axis rotation range of
360$\degree{}$ in one case, and 180$\degree{}$ in another case.  
The purpose of distinguishing 360$\degree{}$- and 180$\degree{}$-rotation
is that when diffraction is present in the sample, the image obtained
from $\theta$ and $\theta + 180\degree{}$ can be different, unlike
conventional tomography. In
order to acquire high-quality results, we conducted a series of
experiments to determine that the optimal values for the regularizer
weights in Eq.~\ref{eqn:loss_func} were
$\alpha_{\delta} = 1.5\times 10^{-8}$ for the stronger phase-shifting
part of the x-ray RI, and
$\alpha_{\beta} = 1.5\times 10^{-9}$ for the weaker absorptive part.
The total variation minimization regularizer term of Eq.~\ref{eqn:tv}
was made small by setting $\gamma=1\times 10^{-11}$.  The x-ray
RI grid was initialized to a Gaussian distribution with
a mean of $\delta=8.7\times 10^{-7}$ and $\beta=5.1\times 10^{-8}$
with standard deviations of about a tenth of the mean.  These values are
lower than the expected values but gave better reconstruction starts
than values of zero; the reconstructions were not sensitive to the
exact non-zero initialization values.  In TensorFlow, we used a
minibatch size of 10, and set the iterator to stop automatically once
the incremental decrease of the total loss function of
Eq.~\ref{eqn:loss_func} fell below 3\%{}.  Parallelized with 4
threads, using 3 levels of multiscaling, and running on CPUs, the
full-field reconstruction finished with 10, 10, and 6 epochs for the
three passes with 4$\times$, 2$\times$, and 1$\times$ (original
resolution) downsampling. The entire computation took approximately
5.0 hours of wall clock time, and 120 core hours.

For ptychography, we assumed that an x-ray optic was used to focus a
beam on the entrance of the object volume with a 
Gaussian profile beam profile
with $\sigma_x = \sigma_y = 6$ nm, and a
maximum probe phase of 0.5 radians.  A total of $23\times 23=529$
probe positions were used to illuminate the specimen from each viewing
angle, as shown in Fig.~\ref{fig:forward}.  The sparsity and
smoothening constraints in ptychography are relaxed by probe overlap,
so that a different set of regularizer values for
Eq.~\ref{eqn:loss_func} were used, with
$\alpha_{\delta} = 1\times 10^{-9}$,
$\alpha_{\beta} = 1\times 10^{-10}$, and $\gamma = 1\times
10^{-9}$. The x-ray RI grid was initialized in the same
way as the full-field case. The minibatch size was set to 1 rather
than 10 so as to allow all data from one projection angle to fit in
computer memory.  For this larger data set with far-field diffraction
intensity recordings, the reconstruction was parallelized with 20
threads, and required 4 epochs to yield a high quality result over a
wall clock time of 46 hours, or 16,500 core hours.

Figure~\ref{fig:cone_results} shows the true object in row (a), and
the reconstruction results for these various approaches in rows (b)
through (e).  In the 360$\degree{}$ full-field reconstruction shown in
row (b), and the 360$\degree{}$ ptychographic reconstruction shown in
row (d), the object boundaries are sharp, and features within the
object are nicely reproduced.  This is decidedly not the case for the
ER+FBP or conventional tomographic reconstruction shown in row (e),
where the small spheres on the outside of the object are poorly
reproduced in the surface view of the fourth column, and the
projection images of columns two and three do not accurately reproduce
the true object.  In visual appearance, one can argue that the
ptychography reconstruction shown in row (d) is slightly sharper and
has least ``ghost'' structure present in what are supposed to be empty
voids inside the cone compared to the full-field reconstruction shown
in row (b).  This may be due to the fact that the large number of
spatially separate illumination patterns used in ptychography help
limit the regions that contribute scattering signal to each of the 529
individual data recordings acquired per rotation angle.  On the other
hand, the ptychographic reconstruction shows some slight fringe
artifacts at the bottom of the cone, which might arise from the fact
that the data are recorded in the far rather than the near field.
More quantitative comparisons will be presented below, using the
Fourier shell correlation (FSC) method
\cite{saxton_jmic_1982,vanheel_jsb_2005} which measures the
consistency between images as a function of spatial frequency
(resolution in the Fourier transform).

\begin{figure}[t]
  \centerline{\includegraphics[width=0.6\textwidth]{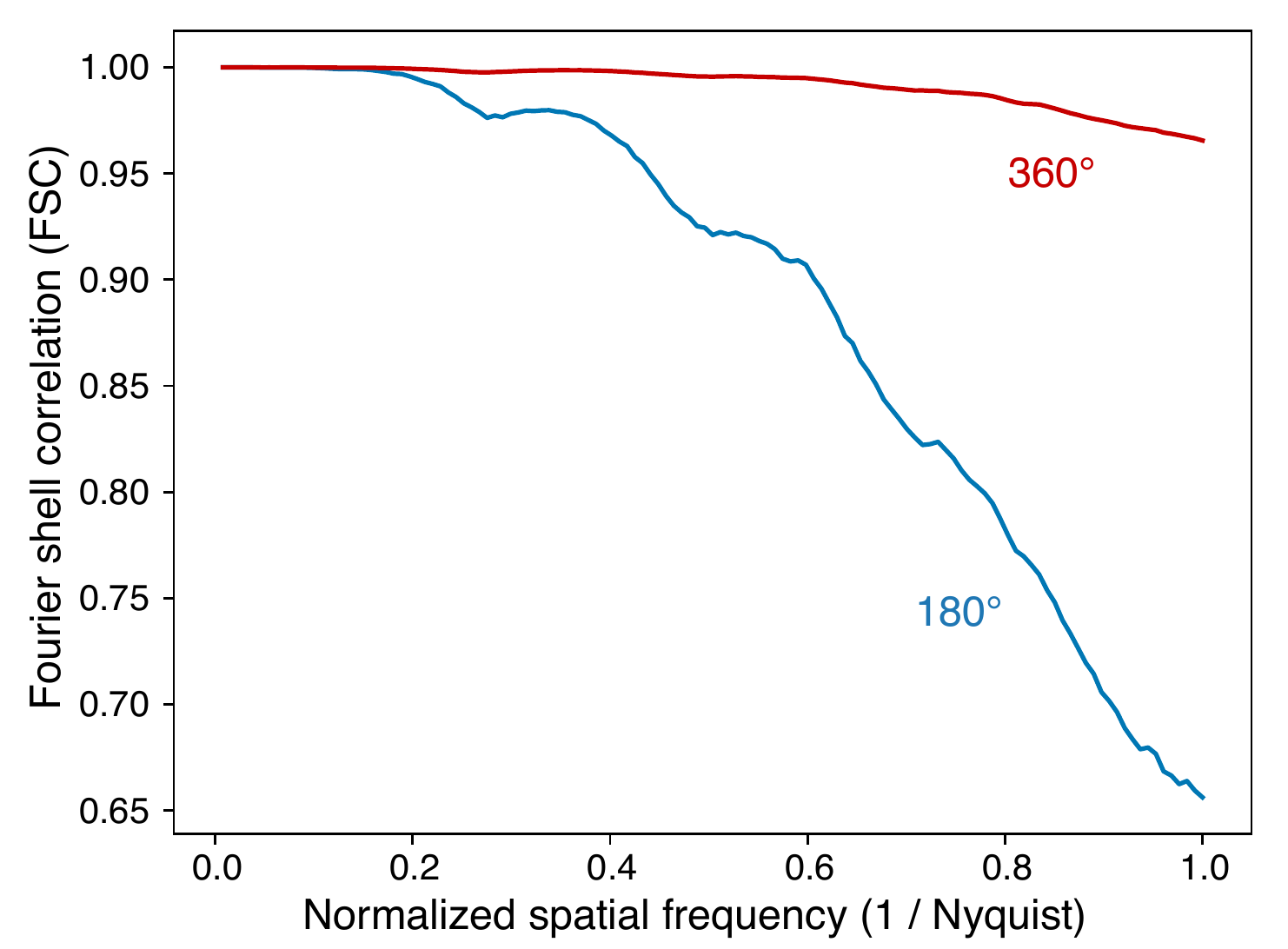}}
  \caption{Fourier shell correlation (FSC) measurement of the
    resolution of full-field reconstructions using 500 rotation angles
    distributed over 360$\degree{}$ [Fig.~\ref{fig:cone_results}(b)] 
    versus only 180$\degree{}$ [Fig.~\ref{fig:cone_results}(c)].  The
    FSC measurement shows a considerable loss in spatial resolution
    over a wide range of spatial frequencies, but especially at the
    highest spatial frequencies corresponding to the finest features.
    The FSC was calculated by comparing the phase shifting part of the
    x-ray RI $\delta$ between 
    two separate iterative reconstructions of the same full-field
    recording dataset.}
  \label{fig:cone_fsc}
\end{figure}

In conventional tomography within the pure projection approximation,
projections obtained 180$\degree{}$ apart are identical after
projection reversal, so that data collected over a 180$\degree{}$
range are sufficient for an accurate reconstruction.  This is not the
case when diffractive effects come into play, as has long been known
in diffraction tomography \cite{devaney_ui_1982} and as was 
observed in our previous study of ptychographic reconstructions of an
object with depth-of-focus effects included
\cite{gilles_optica_2018}. Modulations on the wavefield can be
produced both by Fresnel diffraction from upstream features, and
refractive modulation from features at downstream planes; one cannot
unambiguously distinguish between these two effects using a single
viewing angle.  To illustrate this, we have carried out a simulation
of full-field imaging where the same 500 rotation angles used in the
360$\degree{}$ case were instead distributed over a 180$\degree{}$
angular range, giving the results shown in row (c). This leads
to the presence of a number of voids in the reconstructed refractive
index distribution, presumably because of the ambiguity noted above;
the voids remained in the same position even if the 180$\degree{}$
illumination angles were shifted to a different range, and is not 
related to the shrink-wrap of the finite support.  By removing
the positivity constraint on the RI distribution and
examining the intermediate object function as it was updated after each
minibatch, we noticed that the values in the void regions became
negative and kept decreasing.  We therefore speculate that the voids
might arise as the optimizer attempts to compensate for the loss
function under information deficiency.  The resolution of the
180$\degree{}$ and 360$\degree{}$ full-field reconstructions was
evaluated using the FSC between two independent reconstructions with
the same parameter settings, and the result shown in
Fig.~\ref{fig:cone_fsc} clearly shows the loss of resolution that
results from using the same number of projection angles distributed
over 180$\degree{}$ only.

%

In order to understand the robustness of our reconstruction method in
the presence of noise due to limited exposure, we carried out
full-field reconstructions where the recorded diffraction intensities
were modified to incorporate noise.  (Other studies have considered
the noise robustness of simple coherent diffraction imaging against
zone plate imaging \cite{huang_optexp_2009} or against near-field
imaging \cite{hagemann_jac_2017} with somewhat differing conclusions;
we leave a comparison of full-field imaging and ptychography for
future work).  This was done by setting a quantity $\nph$ to be the
total number of incident photons that intersect the object support
with the object at each of the 500 projection angles.  The object
support is characterized by an area overdetermination ratio (AOR) of
\begin{equation}
 \aor = \frac{N^{2}}{N_{\rm support}}
 \label{eqn:area_oversampling}
\end{equation}
where $N_{\rm support}$ represents the total number of pixels within
the finite support \cite{miao_josaa_1998}.  The Gaussian-blurred
``shrink-wrap'' procedure described above led to $\aor \simeq 57$\%
for the simulated cone object.  With these factors considered, the number of photons
$\npixtheta$ incident on each of the $N^{2}=256^{2}$ pixels
is given by
\begin{equation}
  \npixtheta =
  \frac{\nph}{\aor\times N^{2}}
  \label{eqn:photons_per_pixel}
\end{equation}
for each of the $N_{\theta}$ viewing directions such that
$\nph=1\times 10^{9}$ corresponds to $\npixtheta=2.7 \times 10^{4}$.
The detected intensity images at each of the $N_{\theta}$ angles,
generated by a normalized plane incident wave with unity magnitude,
were scaled by $\npixtheta$ before Poisson noise was applied to them.
We then obtained reconstructions from the Poisson-degraded datasets,
and compared them using the Fourier shell correlation (FSC) method
\cite{vanheel_jsb_2005}.

\begin{figure}[tbp]
  \centerline{\includegraphics[width=0.8\textwidth]{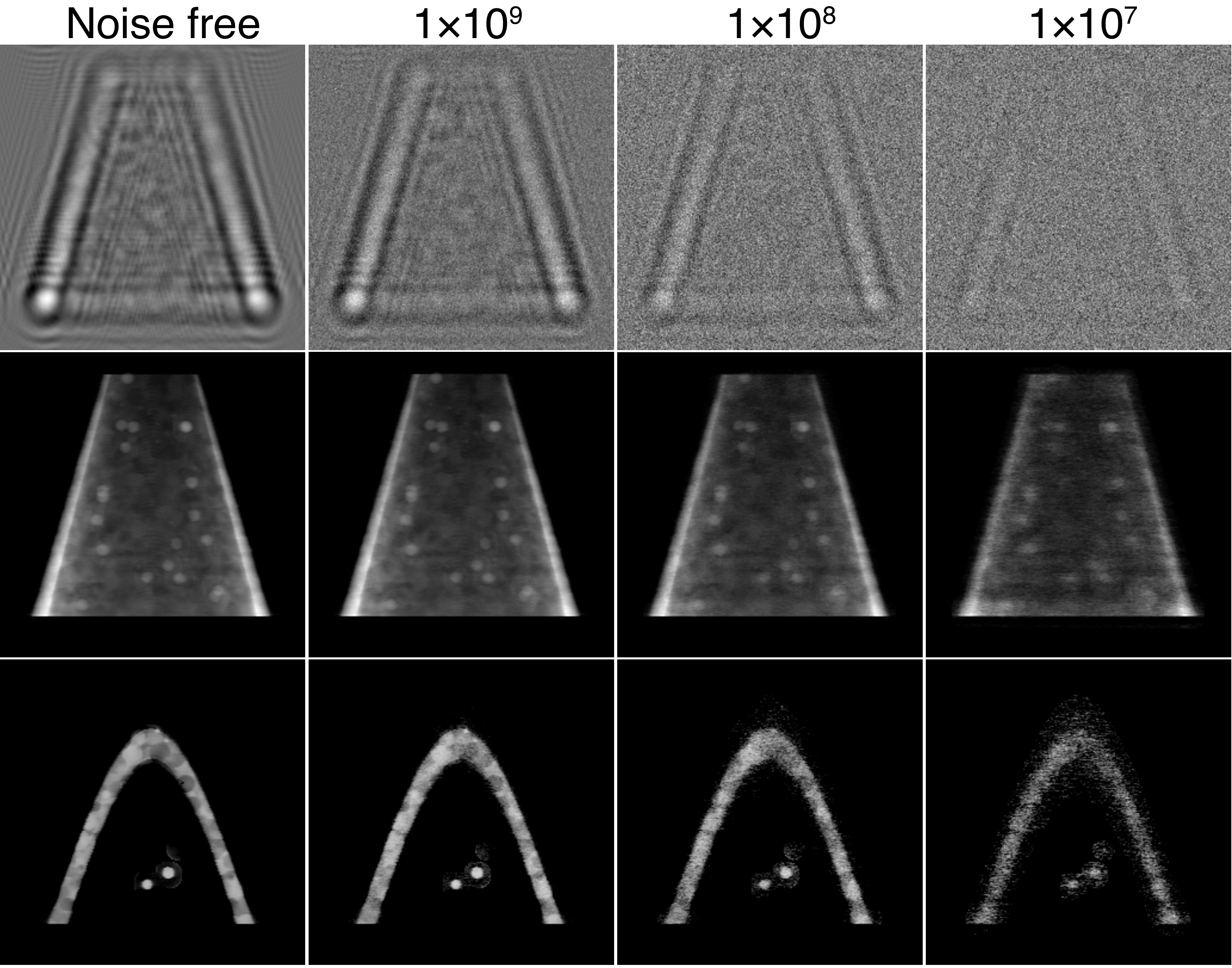}}
  \caption{Near-field propagation images as would be recorded by the
    camera in Fig.~\ref{fig:forward}(a) (top row), and reconstructed
    images (middle and bottom rows) in the presence of simulated
    Poisson noise with a total number of photons $\nph$ used in data
    recording.  The middle row projection images correspond to the
    second column image of row (b) in Fig.~\ref{fig:cone_results},
    while the bottom row slice images correspond to the first column
    image of row (b) in Fig.~\ref{fig:cone_results}.  As can be seen,
    reducing the number of photons $\nph$ within the object (and thus
    the incident number of photons per pixel per viewing angle
    $\npixtheta$ as given by Eq.~\ref{eqn:photons_per_pixel}) leads to
    a decrease in image fidelity and signal to noise ratio at the
    single pixel level.  One can also evaluate this as a loss of
    spatial resolution with decreasing exposure, as shown in
    Fig.~\ref{fig:noise_fsc}.}
  \label{fig:noise_eval}
\end{figure}

\begin{figure}[tbp]
  \centerline{\includegraphics[width=0.5\textwidth]{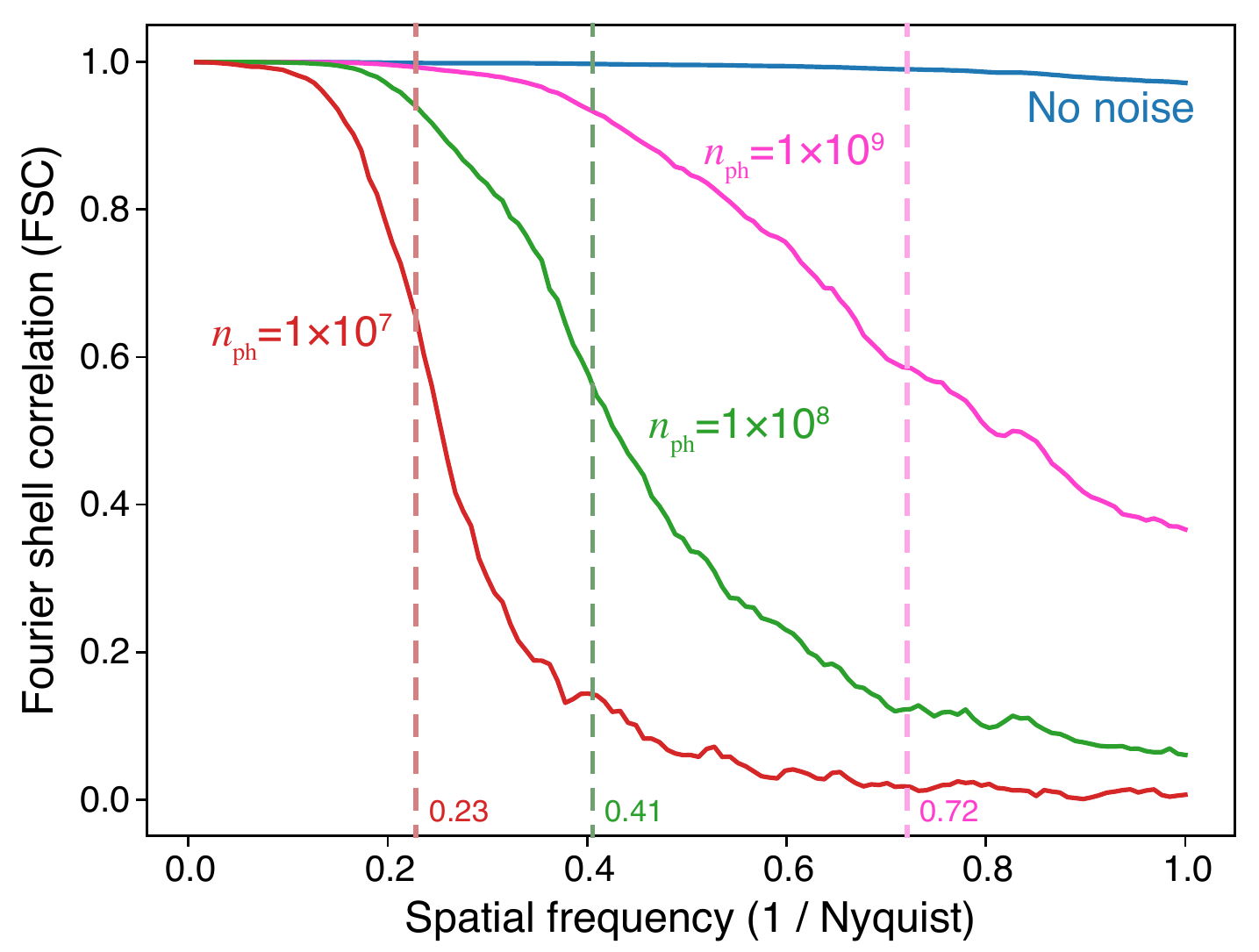}}
  \caption{The Fourier shell correlation (FSC) for the fullfield results
    with varying photon exposure $\nph$ as shown in
    Fig.~\ref{fig:noise_eval}.  For each exposure, two different
    datasets were generated with different Poisson-distributed random noise, after which
    the same reconstruction algorithm with the same parameters was
    applied to each noisy dataset before performing the Fourier shell
    correlation.  As discussed in the text, one would expect the SNR
    to decline at normalized spatial frequencies of
    $\{0.23,0.41,0.72\}$ times the normalized spatial frequency for
    photon exposures of $\nph=\{1\times 10^{7},1\times 10^{8},1\times
    10^{9}\}$.  The dashed lines at these normalized spatial
    frequencies are all at roughly consistent decreases of the FSC to
    about 0.5--0.6 for the respective exposures, corresponding to a
    spatial resolution estimate of $\{4.4,2.5,1.4\}$ nm.}
  \label{fig:noise_fsc}
\end{figure}

Given the normalized image intensity one would expect from a
feature-present versus a feature-absent voxel, one can estimate the
exposure required to see that object with a specified signal-to-noise
ratio (SNR).  Using the phase contrast imaging expression of Eq.~39 of
\cite{du_ultramic_2018} for $t=1$ nm thick Si at 5 keV, we obtain an
exposure estimate of $\npixtheta=5.0\times 10^{7}$ for SNR=5 imaging.
Dose fractionation \cite{hegerl_zn_1976} tells us that this dose can
be distributed over all $N_{\theta}$ viewing angles as 3D object
statistics are built up from tomographic projections, so one would
expect that to achieve full resolution at SNR=5 one would require
$\npixtheta/N_{\theta}=1.0\times 10^{5}$, which from
Eq.~\ref{eqn:photons_per_pixel} translates to $\nph=3.7\times 10^{9}$
per viewing angle.  Because this exposure scales as SNR$^{2}$,
reducing $\nph$ from $3.7\times 10^{9}$ to $1.0\times 10^{9}$
corresponds to a decrease of SNR from 5 to $5/\sqrt{3.7/1.0}=2.6$.
Alternatively, because the radiation dose that must be necessarily
imparted to a specimen for imaging at a specified SNR scales as the
inverse fourth power of spatial resolution
\cite{howells_jesrp_2009,du_ultramic_2018}, a decrease of $\nph$ from
$3.7 \times 10^{9}$ to
$\{1\times 10^{7},1\times 10^{8},1\times 1^{9}\}$ would be expected to
correspond to a reduction in spatial resolution from 1 to
$\{4.4,2.5,1.4\}$ nm.  If one translates this into a fraction of the
Nyquist sampling limit, the corresponding fractions are
$\{0.23,0.41,0.72\}$.  These fractions of the Nyquist sampling limit
are shown via dashed lines in Fig.~\ref{fig:noise_fsc}, and they are
consistent with a FSC in the 0.5--0.6 range as a measure of the
spatial resolution limit.

\section{Discussion}

We have shown here an approach (which we call Adorym, as noted above)
whereby one can use automatic differentiation and a multislice
propagation forward model to reconstruct 3D
objects in two different microscope types, with only minor
branchpoints in one computer code base.  Our approach has the
following characteristics which are shared with another related
non-automatic-differentiation approach \cite{gilles_optica_2018}, as
well as with other multislice learning \cite{kamilov_optica_2015} and
optimization-based \cite{kamilov_ieeetci_2016} approaches:
\begin{itemize}

\item In standard diffraction tomography approaches, one assumes that
  the 3D object can be decomposed into volume gratings so that data
  from one viewing angle is projected not onto a flat plane in Fourier
  space (as would be the case for a pure projection image), but the
  surface of the Ewald sphere.  These assumptions are valid for the
  case where there is no multiple scattering in the specimen, but it
  can be shown (see for example Figs.~2 and 3 in
  \cite{du_ultramic_2018}) that multiple scattering can play a role in
  x-ray imaging of thick specimens.  In other words, the illumination
  of downstream planes can be affected by the presence of strong
  features in upstream planes. Because our approach involves a
  full multislice forward calculation along each viewing angle, it can
  incorporate these effects correctly.

\item In previous multislice ptychography approaches
  \cite{maiden_josaa_2012}, it has been assumed that the object can be
  decomposed into a set $N_{a}$ of planes along the illumination
  direction with those planes separated by a depth of focus distance
  or more.  This separation is required for allowing the combination
  of propagated probe, and discrete object plane, to be decomposed
  into into sufficiently different results along the propagation
  direction.  By allowing for an isotropic forward model where the
  plane separation distance can be the same as the transverse
  resolution distance, our approach is better able to represent the
  subtle contrast variations that occur in imaging over distances that
  are a reasonable fraction of the depth of focus.  It should also be
  noted that the use of multislice methods to reconstruct $N_{a}$
  planes along the illumination direction means one can reduce the
  number of illumination angles used \cite{jacobsen_optlett_2018} from
  what one might have expected based on the Crowther criterion
  \cite{crowther_prsa_1970}.

\item In previous ptychographic tomography \cite{dierolf_nature_2010}  and multislice
  ptychographic tomography approaches \cite{li_scirep_2018}, phase
  unwrapping methods have been used to generate a pure projection
  image with a linear response. No phase unwrapping is required in our
  approach, eliminating any potential errors and ambiguities that can
  sometimes occur with phase unwrapping.

\end{itemize}
One would expect these characteristics to allow for the reconstruction
of beyond-depth-of-focus 3D objects with greater fidelity than one
would have with multislice ptychographic tomography approaches;
exploration of this hypothesis could be the topic of future work.

Our use of automatic differentiation in a numerical optimization
approach has several features:
\begin{itemize}

\item For ptychography, it lets one phase the far-field Fourier
  magnitudes as was first suggested \cite{jurling_josaa_2014} and then
  demonstrated \cite{nashed_procedia_2017} in prior work.  For
  near-field imaging, it avoids the approximations of uniform material
  type implicit in one commonly-employed approach
  \cite{paganin_jmic_2002}.

\item It allows one to easily switch between different imaging modes
  (in this example, both near-field imaging and ptychography) within
  the same code framework, and it lets one explore different types of loss
  functions and regularizers without needing to rebuild the optimzer.
  This can be very useful for benchmarking different imaging and 
  reconstruction techniques. 

\item Several software packages that provide automatic differentiation
  capabilities (such as TensorFlow and Autograd) are already built for
  parallelized operation on large compute clusters.  As an example, an
  automated diffrentiation based ptychography reconstruction code was
  demonstrated in \cite{nashed_procedia_2017}. 
  We carried out a direct test of our TensorFlow based AD approach for
  ptychography against our previous result \cite{gilles_optica_2018}
  using manual differentiation of the cost function, and
  implementation in C++.  The approach used here took 8.25 core
  hours/iteration/angle, compared to 6.48 core
  hours/iteration/angle. One pays a modest penalty in computer time in
  this example, but arguably uses less researcher time because
  automatic differentiation does not require one to re-calculate
  derivatives as the cost function is modified.
  
\item It also allows one to trivially compare and switch between synchronous and asynchronous
schemes of optimization. In the synchronous scheme, object functions are broadcasted and
synchronized among all threads for each several iterations. In the asynchronous scheme,
each threads does the optimization own its own, and the object functions contained by them
are only combined at the end. In the example discussed in \cite{nashed_procedia_2017}, 
the synchronous approach took slightly longer to complete but gave more accurate
results. The default scheme that we used to generate the above
demonstrative data is a variant of the synchronous approach. Here, each thread keeps its own object
function, but the gradient obtained by a thread is broadcasted and averaged along with
the results of all other threads before being used to update the object function. 

\end{itemize}
The above stated characteristics have led to increasing attention to automatic differentiation
in the optics community, and other work has explored the use of automatic differentiation for
several other coherent diffraction imaging modalities \cite{kandel_optexp_2019}.

Another point needing attention is that while the full-field mode and the
ptychography mode are different in terms of acquisition method and processing
wall time, the results they give are sometimes not equivalent as well. This is
most obvious for diffusive features without clear boundaries, as in the case
demonstrated in Fig.~S1 in the supplementary material. In this test case,
a protein molecule (originally acquired using electron microscopy tomography)
was numerically reconstructed by our algorithm using both full-field mode
and ptychography mode. The result shows that the full-field reconstruction
``throws away'' the diffusive halo around the molecule, which on the other hand
is correctly restored by ptychography reconstruction.

While the proposed method has been implemented for both full-field microscopy
and for ptychography, a special note should be given to the former. 
In the absence
of the oversampling constraint in ptychography, we explored the use of
finite support constraint and sparsity constraint in 3D space, which would
provide more insights to the iterative retrieval of a bounded 3D object by
solving an underdetermined system. The algorithm, when combined with non-scanning
high-resolution imaging techniques, can potentially become the launchpad
for a high-throughput imaging pipeline for measuring thick samples with sub-100-nm
resolution. One of such possible paths is to apply the algorithm to point-projection
microscopy \cite{nixon_prsla_1955,sowa_optica_2018}. While far-field
diffraction patterns suffer a loss of speckle contrast as one goes
from fully coherent to decreasingly partially coherent illumination
(with the best results obtained when the coherence width of the beam
equals the size of the object array \cite{spence_ultramic_2004}), with
near-field wave propagation one only needs to have the spatial
coherence match the distance $\lambda z/(\Delta x)$ over which one has
the ability to record near-field fringes.  Thus point-projection
near-field imaging is able to make use of a greater fraction of
partially coherent sources, such as today's synchrotron light
sources.  At the same time, if one does have full spatial coherence,
the separation of subregions of the object into separate experimental
recordings (diffraction patterns from limited-size illumination spots)
gives reconstructions with better fidelity even with a more relaxed
imposition of regularizers. \textcolor{black}{Moreover, application of the proposed
method to detection methods beyond X rays -- for example, broadband
radiation used for atmospheric transmission \cite{juan_opteng_1999}
and seismology \cite{devaney_ieee_1984}, might also be exploited to
better model the dynamic diffraction of waves propagating in complicated
media over long distances. }



\section{Conclusion}

We have developed and demonstrated a novel 3D reconstruction algorithm
for objects beyond the depth-of-focus limit. The algorithm uses multislice propagation
as the forward model, and retrieve the RI map of the object by minimizing
a loss function containing the squared difference between the amplitudes of the 
forward-propagated wavefront and the measured signals at all viewing angles. 
We implemented the algorithm for both full-field and ptychography imaging, and compared
them in terms of computation walltime and reconstruction fidelity. 
Investigation on the full-field version allowed us to explore 
the constraint requirements for reconstructing of bounded 3D objects with non-scanning
imaging techniques, where sparsity and finite-support constraints are used to
ensure a successful reconstruction. Another novelty of our method lies on the use
of automatic differentiation, which not only prevents the laborious manual differentiation
involved in numerical optimization, but also makes the computational model extremely
adaptable and flexible. Numerical studies of the algorithm using simulated objects
indicate that the proposed method is capable of recovering a thick object with high
spatial resolution and good accuracy. Further validation of the approach with 
experimental data will be our subsequent step, which would examine the capability
of the method in handling noises and help us identify practical challenges such
as probe alignment \cite{guizar_optexp_2008}. The ultimate goal is to combine the
algorithm with high-resolution imaging techniques, which, for full-field imaging, could be 
the point-projection microscopy. The combination of the two could potentially lead to the 
development of a high-throughput pipeline for imaging thick samples.

\section{Acknowledgement}

This research used resources of the Advanced Photon Source and the
Argonne Leadership Computing Facility, which are U.S. Department of
Energy (DOE) Office of Science User Facilities operated for the DOE
Office of Science by Argonne National Laboratory under Contract
No. DE-AC02-06CH11357. We thank the National Institute of Mental
Health, National Institutes of Health, for support under grant R01 
MH115265.

\bibliographystyle{Science}
\bibliography{mybib}

\end{document}